\documentclass[aps,prl,twocolumn,showpacs,superscriptaddress,nofootinbib]{revtex4-2}  
\usepackage[english]{babel}
\usepackage{graphicx}  
\usepackage{dcolumn}   
\usepackage{bm}        
\usepackage{amsmath,amssymb}   
\usepackage{ifpdf,latexsym,overpic,rotating}
\usepackage{multirow}
\usepackage{epsfig}
\usepackage{amsmath}
\usepackage{color}
\usepackage{hyperref}
\usepackage[capitalise]{cleveref} 

\begin{document}

%

\newcommand{\HISKP}{Helmholtz-Institut f\"ur Strahlen- und Kernphysik, University of Bonn, D-53115 Bonn, Germany}
\newcommand{\YORK}{Department of Physics, University of York, Heslington,  York, Y010 5DD, UK}
\newcommand{\REGINA}{University of Regina, Regina, SK S4S0A2 Canada}
\newcommand{\BASEL}{Department  of  Physics,  University  of  Basel,  Ch-4056  Basel,  Switzerland}
\newcommand{\MAINZ}{Institut f\"ur Kernphysik, University of Mainz, D-55099  Mainz, Germany}
\newcommand{\KENT}{Kent State University, Kent, Ohio 44242, USA}
\newcommand{\GLASGOW}{SUPA School of Physics and Astronomy, University of Glasgow, Glasgow, G12 8QQ, UK}
\newcommand{\DUBNA}{Joint Institute for Nuclear Research, 141980 Dubna, Russia}
\newcommand{\PAVIAINFN}{INFN Sezione di Pavia, I-27100 Pavia, Pavia, Italy}
\newcommand{\WASHINGTON}{Center for Nuclear Studies, The George Washington  University, Washington, DC 20052, USA}
\newcommand{\HALIFAXCANADA}{Department of Astronomy and Physics, Saint Mary’s  University, E4L1E6 Halifax, Canada}
\newcommand{\PAVIA}{Dipartimento di Fisica, Universit\`a di Pavia, I-27100  Pavia, Italy}
\newcommand{\GIESSEN}{II. Physikalisches Institut, University of Giessen,  D-35392 Giessen, Germany}
\newcommand{\BOCHUM}{Institut f\"ur Experimentalphysik, Ruhr Universit\"at, 44780 Bochum, Germany}
\newcommand{\MOSCOW}{Institute for Nuclear Research, RU-125047 Moscow,  Russia}
\newcommand{\MOUNTALLISON}{Mount Allison University, Sackville, New Brunswick E4L1E6, Canada}
\newcommand{\ZAGREB}{Rudjer Boskovic Institute, HR-10000 Zagreb, Croatia}
\newcommand{\MASSACHUSETTS}{University of Massachusetts, Amherst, Massachusetts 01003, USA}
\newcommand{\LA}{University of California Los Angeles, Los Angeles, California 90095-1547, USA}
\newcommand{\JERUSALEM}{Racah Institute of Physics, Hebrew University of Jerusalem,  Jerusalem 91904, Israel}
\newcommand{\NEWJERSEY}{Department of Physics and Astronomy, Rutgers University, Piscataway, New Jersey, 08854-8019}
\newcommand{\JLAB}{Thomas Jefferson National Accelerator Facility, Newport News, VA 23606, USA}

\title{First measurement using elliptically polarized photons of the double-polarization observable $E$ for $\gamma p \to p \pi^0$ and $\gamma p \to n \pi^+$}

%
\affiliation{\HISKP}
\affiliation{\GLASGOW}
\affiliation{\BASEL}
\affiliation{\MAINZ}
\affiliation{\REGINA}
\affiliation{\YORK}
\affiliation{\KENT}
\affiliation{\DUBNA}
\affiliation{\PAVIAINFN}
\affiliation{\WASHINGTON}
\affiliation{\HALIFAXCANADA}
\affiliation{\PAVIA}
\affiliation{\GIESSEN}
\affiliation{\BOCHUM}
\affiliation{\MOSCOW}
\affiliation{\MOUNTALLISON}
\affiliation{\ZAGREB}
\affiliation{\MASSACHUSETTS}

\author{F.~Afzal}\email[Corresponding author: ]{afzal@hiskp.uni-bonn.de}\affiliation{\HISKP}
\author{K.~Spieker}\affiliation{\HISKP}
\author{P.~Hurck}\affiliation{\GLASGOW}
\author{S. Abt}\affiliation{\BASEL}
\author{P. Achenbach}\affiliation{\MAINZ}
\author{P. Adlarson}\affiliation{\MAINZ}
\author{Z. Ahmed}\affiliation{\REGINA}
\author{C.S. Akondi}\affiliation{\KENT}
\author{J.R.M. Annand}\affiliation{\GLASGOW}
\author{H.J. Arends}\affiliation{\MAINZ}
\author{M. Bashkanov}\affiliation{\YORK}
\author{R. Beck}\affiliation{\HISKP}
\author{M. Biroth}\affiliation{\MAINZ}
\author{N. Borisov}\affiliation{\DUBNA}
\author{A. Braghieri}\affiliation{\PAVIAINFN}
\author{W.J. Briscoe}\affiliation{\WASHINGTON}
\author{F. Cividini}\affiliation{\MAINZ}
\author{C. Collicott}\affiliation{\HALIFAXCANADA}
\author{S. Costanza}\affiliation{\PAVIA}\affiliation{\PAVIAINFN}
\author{A. Denig}\affiliation{\MAINZ}
\author{M. Dieterle}\affiliation{\BASEL}
\author{E.J. Downie}\affiliation{\WASHINGTON}
\author{P. Drexler}\affiliation{\MAINZ}\affiliation{\GIESSEN}
\author{S. Fegan}\affiliation{\YORK}
\author{S. Gardner}\affiliation{\GLASGOW}
\author{D. Ghosal}\affiliation{\BASEL}
\author{D.I. Glazier}\affiliation{\GLASGOW}
\author{I. Gorodnov}\affiliation{\DUBNA}
\author{W. Gradl}\affiliation{\MAINZ}
\author{D. Gurevich}\affiliation{\MOSCOW}
\author{L. Heijkenskj\"old}\affiliation{\MAINZ}
\author{D. Hornidge}\affiliation{\MOUNTALLISON}
\author{G.M. Huber}\affiliation{\REGINA}
\author{V.L. Kashevarov}\affiliation{\MAINZ}\affiliation{\DUBNA}
\author{S.J.D. Kay}\affiliation{\YORK}
\author{M. Korolija}\affiliation{\ZAGREB}
\author{B. Krusche}\affiliation{\BASEL}
\author{A. Lazarev}\affiliation{\DUBNA}
\author{K. Livingston}\affiliation{\GLASGOW}
\author{S. Lutterer}\affiliation{\BASEL}
\author{I.J.D. MacGregor}\affiliation{\GLASGOW}
\author{R.G. Macrae}\affiliation{\GLASGOW}
\author{D.M. Manley}\affiliation{\KENT}
\author{P.P. Martel}\affiliation{\MAINZ}\affiliation{\MOUNTALLISON}
\author{R. Miskimen}\affiliation{\MASSACHUSETTS}
\author{M. Mocanu}\affiliation{\YORK}
\author{E. Mornacchi}\affiliation{\MAINZ}
\author{C. Mullen}\affiliation{\GLASGOW}
\author{A. Neganov}\affiliation{\DUBNA}
\author{A. Neiser}\affiliation{\MAINZ}
\author{M. Oberle}\affiliation{\BASEL}
\author{M. Ostrick}\affiliation{\MAINZ}
\author{P.B. Otte}\affiliation{\MAINZ}
\author{D. Paudyal}\affiliation{\REGINA}
\author{P. Pedroni}\affiliation{\PAVIAINFN}
\author{A. Powell}\affiliation{\GLASGOW}
\author{G. Reicherz}\affiliation{\BOCHUM}
\author{T. Rostomyan}\affiliation{\BASEL}
\author{C. Sfienti}\affiliation{\MAINZ}
\author{V. Sokhoyan}\affiliation{\MAINZ}
\author{O. Steffen}\affiliation{\MAINZ}
\author{I.I. Strakovsky}\affiliation{\WASHINGTON}
\author{T. Strub}\affiliation{\BASEL}
\author{I. Supek}\affiliation{\ZAGREB}
\author{A. Thiel}\affiliation{\HISKP}
\author{M. Thiel}\affiliation{\MAINZ}
\author{A. Thomas}\affiliation{\MAINZ}
\author{Yu.A. Usov}\affiliation{\DUBNA}
\author{S. Wagner}\affiliation{\MAINZ}
\author{N.K. Walford}\affiliation{\BASEL}
\author{D.P. Watts}\affiliation{\YORK}
\author{D. Werthm\"uller}\affiliation{\YORK}
\author{J. Wettig}\affiliation{\MAINZ}
\author{L. Witthauer}\affiliation{\BASEL}
\author{M. Wolfes}\affiliation{\MAINZ}
\author{N. Zachariou}\affiliation{\YORK}

\collaboration{A2 Collaboration at MAMI}
 \vskip 0.25cm
\date{\today}

\begin{abstract}
We report the measurement of the helicity asymmetry $E$ for the $p\pi^0$ and $n\pi^+$ final states using, for the first time, an elliptically polarized photon beam in combination with a longitudinally polarized target at the Crystal Ball experiment at MAMI. The results agree very well with data that were taken with a circularly polarized photon beam, showing that it is possible to simultaneously measure polarization observables that require linearly (e.g.~$G$) and circularly polarized photons (e.g.~$E$) and a longitudinally polarized target. The new data cover a photon energy range 270 - 1400 MeV for the $p\pi^0$ final state (230 - 842 MeV for the $n\pi^+$ final state) and the full range of pion polar angles, $\theta$, providing the most precise measurement of the observable $E$. A moment analysis gives a clear observation of the $p\eta$ cusp in the $p\pi^0$ final state.
\end{abstract}

\maketitle
The dynamics of quarks and gluons within nucleons can be studied through the excitation spectrum they form as composite systems. Studying the masses, widths and decays of the light baryon resonances provides important information about the strong interaction in the non-perturbative regime of quantum chromodynamics (QCD). Quark model calculations based on three constituent quarks \cite{Isgur:1978xj,CAPSTICK2000S241,Loring:2001kx} predict a large number of states, which is qualitatively confirmed by Lattice QCD calculations albeit quark mass parameters with a resulting pion mass of $m_\pi=396$~MeV are used \cite{Edwards:2011jj}. Additionally, hybrid baryons are predicted as well, increasing the number of expected states \cite{Dudek:2012ag}. However, the predicted dense light baryon spectrum is not observed in experimental data. An overview of the intensive experimental efforts in studying the properties of light baryon resonances is given in \cite{Thiel:2022xtb}. At the Crystal Ball experiment located at the Mainz Microtron (MAMI) the nucleon excitation spectrum is probed with a real photon beam, allowing several different photoproduction reactions to be studied. In the case of single pseudoscalar meson photoproduction (e.g. $\gamma p\to p\pi^0$ and $\gamma p \to n\pi^+$), the magnitudes and phases of four complex amplitudes have to be determined uniquely. In addition to measuring the unpolarized cross section, several single- and double-polarization observables must be measured for a ``complete" analysis \cite{ChiangTabakin,Wunderlich20}, employing a polarized photon beam, polarized target or recoil nucleon polarimetry \cite{Sandorfi}. The resonance parameters are extracted from the data using different partial wave analysis (PWA) approaches, e.g. the BnGa \cite{BnGa2017} and SAID \cite{SAIDMA19} groups use a K-matrix approach, while the JüBo-model \cite{JuBo17} uses a dynamically coupled-channel approach.\\
In the last decade, several single and double polarization observables have been measured for the $p\pi^0$ final state and their impact on the extracted partial waves were studied by calculating the variances between partial waves from different PWA approaches before and after including the polarization observables in their respective PWAs (see Fig.~9 in \cite{Anisovich:2016vzt}). The measured polarization observables significantly reduced the variances between the different approaches. A large contribution to the previously existing discrepancies were attributed to the $p\pi^0$ $S$ wave. Despite the new data and the overall reduced variances between the PWAs, significant deviations remain at low energies ($E_\gamma<$ $800$~MeV), in particular directly at the $p\eta$ photoproduction threshold (see Fig.~9 in \cite{Anisovich:2016vzt}), where a cusp effect in the $S$ wave is visible. So far, the $p\eta$ cusp was observed in the unpolarized cross section data of the $p\pi^0$ and $n\pi^+$ final states \cite{Althoff,Adlarson} and helicity dependent cross sections of the $n\pi^+$ final state \cite{PhysRevC.74.045204}. The correct implementation of all singularities, e.g.~resonance poles and branch points, is crucial for the correct extraction of partial waves and resonance parameters \cite{Ceci,Afzal20}. The double polarization observable $E$, which requires a circularly polarized photon beam and a longitudinally polarized target, has a sensitivity to the $S$ wave due to an interference with e.g. $D$ waves. However, previous measurements of $E$ from CBELSA/TAPS \cite{Epi0prl, Epi0long} and CLAS \cite{CLAS:2023ddn} lack full angular coverage and high precision at low energies.

At low energies, the circular polarization degree $p_\gamma^{\text{circ}}$ is low because it increases with the photon energy $E_\gamma$ according to Olsen \cite{olsen}:
\begin{equation}
\label{eq:circpol}
p_\gamma^{\text{circ}}=p_e·(4x-x^2)/(4-4x+ 3x^2),
\end{equation}
which was derived for an amorphous radiator with $x=E_\gamma/E_{e^-}$, $p_e$ being the electron polarization degree and $E_{e^-}$ being the incident electron beam energy. To obtain high-precision data with a larger photon flux at low energies, the Crystal Ball experiment at MAMI employed for the first time elliptically polarized photons. This is achieved by combining longitudinally polarized electrons with a crystal radiator that leads to elliptically polarized photons for the coherent bremsstrahlung part, which has both a linear and a circular polarization component, and to circularly polarized photons for the incoherent bremsstrahlung part according to Eq.~\eqref{eq:circpol}. For the coherent bremsstrahlung part, sharp peaks (known as coherent peaks) arise in the intensity spectrum for certain crystal orientations caused by the periodical structure of the crystal lattice. First theoretical calculations of the circular polarization degree \cite{Bosted2001} predict a small dependence on the crystal lattice which appears as small dips in the curve of the circular polarization degree at the coherent edge positions, slightly reducing the circular polarization degree. However, no theoretical calculations are reported for $x<0.5$ and no experimental data are available presently to verify these calculations. \newline
We report here the first data that were taken with elliptically polarized photons together with a longitudinally polarized target at the Crystal Ball experiment at MAMI, allowing the double-polarization observables $E$ (circ.~pol.) and $G$ (lin.~pol.) to be measured simultaneously as the polarized cross section for this configuration reads \cite{Sandorfi} 

\begin{equation}
\begin{split}
\label{equ:polcs_EG}
\frac{d\sigma}{d\Omega}_{\text{pol}}(E_\gamma,\cos\theta,\varphi) = \frac{d\sigma}{d\Omega}_{0}(E_\gamma,\cos\theta)\times \\ \Big[ 1-  p_\gamma^{\text{lin}} \Sigma  \cos(2\varphi) + p_T p_\gamma^{\text{lin}} G \sin(2\varphi) - p_T p_\gamma^{\text{circ}} E \Big].
\end{split}
\end{equation} 
$\frac{d\sigma}{d\Omega}_{0}(E_\gamma,\cos\theta)$ is the unpolarized cross section, the angle $\varphi$ is the angle between the linear polarization vector and the reaction plane, $p_\gamma^{\text{lin}}$ and $p_\gamma^{\text{circ}}$ are the degrees of linear and circular polarization, respectively, and $p_T$ is the degree of target polarization. The results for the doule polarization observable $G$ are shown in the Supplemental Material \cite{afzal23suppl}.
In this Letter, we focus on the results of the double-polarization observable $E$ for the $p\pi^0$ and the $n\pi^+$ final states and demonstrate the main advantage of measuring $E$ with elliptically polarized photons: the availability of a higher photon flux, as well as the obvious efficiency of simultaneously measuring more than one observable. In particular, the higher photon flux due to the coherent bremsstrahlung leads to high-precision data that has allowed effects such as the $p\eta$ cusp to be observed in a double-polarization observable. \newline

\begin{figure}[h]
\includegraphics[width=\columnwidth]{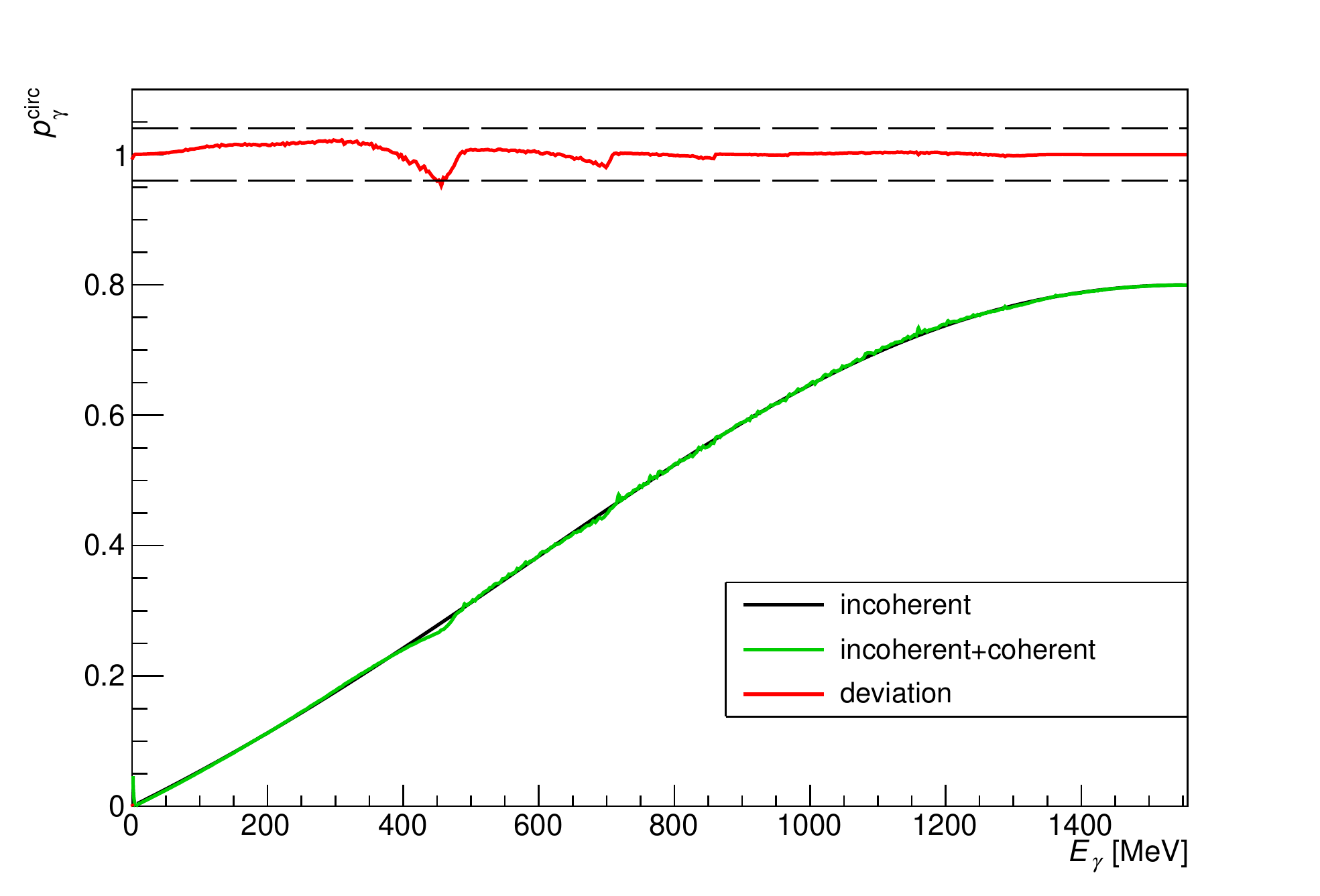}
\caption{The degree of circular polarization of the bremsstrahlung photons are calculated for only the incoherent part (black) according to Eq.~\eqref{eq:circpol} with $p_e= 80$\% and also when including the coherent part (green) for a 450 MeV coherent edge setting for the diamond as used for the data shown in Fig.~\ref{fig:DiaMoe}. The ratio between (incoherent+coherent) and incoherent (red) shows a deviation smaller than 4\% (dashed lines) at all photon energies.}
\label{fig:pol}
\end{figure}

The data were collected in four separate beamtimes at the Crystal Ball experiment located at the MAMI accelerator in Mainz, Germany, from November 2013 until September 2015. A longitudinally polarized electron beam with an energy of $E_{e^-}=1557$ MeV was provided by MAMI \cite{mami2}, whereby the helicity of the electrons was switched with a frequency of 1 Hz. The polarization degree of the electrons $p_e$ was measured daily using a Mott polarimeter \cite{mott}. The electron beam was incident on a thin diamond crystal radiator, providing elliptically polarized photons via bremsstrahlung. Different coherent edge positions were chosen from 350 - 850 MeV in 100 MeV steps during the measurement with the diamond radiator, giving a maximum photon linear polarization of between 75\% and 45\% near the various coherent edge positions. See \cite{Lohmann} for a detailed explanation of this process. 
In addition, measurements were performed in regular time intervals using an amorphous foil radiator of 10 $\mu$m thickness \cite{Moeller}, which produced only a circularly polarized photon beam. This allowed the electron polarization degree to be cross checked, using a M{\o}ller polarimeter \cite{Moeller}, against the results from the Mott polarimeter. Both measurements were in good agreement with an average of $p_e\approx 80$\%. The degree of circular polarization $p_\gamma^{\text{circ}}$ was determined according to Eq.~\eqref{eq:circpol} (see Fig.~\ref{fig:pol}). A first calculation of the circular polarization degree including the coherent photons and using the diamond orientation of the 450 MeV setting is also shown as comparison. Similar to the calculations from Bosted \cite{Bosted2001}, we observe dips in the circular polarization degree at the positions of the main coherent edge position of 450 MeV ($x\approx 0.3$), corresponding to the reciprocal lattice vector [022], as well as at the position other higher order reciprocal lattice vectors which contribute at higher photon energies (see ehancement spectrum in Fig.~1 in the Supplemental Material~\cite{afzal23suppl}). The differences in the circular polarization degree between the incoherent and (incoherent+coherent) bremsstrahlung photons are very small ($<4$\%). The code used for the calculation is based on \cite{Jones:2023} and further details about the calculation are discussed in the Supplemental Material \cite{afzal23suppl} and elsewhere \cite{Afzal:2023}.

The Glasgow-Mainz tagging system \cite{tagger3} was used to energy-tag the photons that were incident on a longitudinally polarized frozen-spin butanol Mainz-Dubna target \cite{target1} of 2 cm length. A maximum target polarization degree of $p_T=89$\% was achieved with relaxation times of more than 1000 h. 

The final state particles were detected mainly by two calorimeters: the Crystal Ball (CB) \cite{cb} and the TAPS \cite{taps1} detectors. Together, both calorimeters covered the polar angle $\theta$ range from $5^\circ$ to $160^\circ$ and the entire azimuthal angle $\phi$ range ($2\pi$ radians). In addition, charged particles were identified by the PID-II detector, the Multi-Wire Proportional Chambers and plastic scintillators that were placed in front of the TAPS crystals. More details regarding the experimental setup are given in \citep{Adlarson,Werthmueller14,Dieterle2018}.

For the $p\pi^0 \to p \gamma \gamma$ final state, two- or three-cluster events in both calorimeters together were retained for the analysis. In case of a three-cluster event, all possible combinations were used to calculate the two photon invariant mass and a $2\sigma$ cut was utilized to select the $\pi^0$ meson. Events with exactly two clusters in the calorimeters were used by assigning the two clusters to the two decay photons of the $\pi^0$. In contrast, exactly one charged and one uncharged cluster were demanded for the $n\pi^+$ final state. If the neutron was lost, events with exactly one charged cluster were analyzed as well. To reject background events, cuts on several standard kinematic parameters were applied for both final states, i.e. the missing mass, the coplanarity and the pulse-shape analysis, which are described for a similar analysis in \citep{Werthmueller14,Dieterle2018}. Further details of the analysis can be found in \citep{PhDthesis,PhDKarsten}.

To determine the amount of unpolarized background contributions from carbon and helium nuclei present in the frozen spin butanol target, data were also taken with a carbon foam target that was cooled down with the same mixture of $^3$He/$^4$He as for the butanol target. The carbon+helium data were scaled with a factor $s$ to the butanol data using e.g.~the coplanarity spectra in a range where the free nucleons do not contribute. The proportion of free protons in the selected data is then described by the dilution factor $d$:
\begin{equation}
\label{equ:dilution}
d= \frac{N^{\text{free}}}{N^{\text{free}}+N^{\text{bound}}}= 1- s \frac{N^{\text{bound}}}{N^{\text{free}}+N^{\text{bound}}},
\end{equation}
where $N^{\text{free}}$ and $N^{\text{bound}}$ are the event yields for the free and bound nuclei.
 
The double-polarization observable $E$ was determined separately from $\Sigma$ and $G$. Integrating over the full azimuthal angular range and noting that the sign of the linear polarization does not change when the electron helicity is flipped, the contributions from $\Sigma$ and $G$ cancel and the double-polarization observable $E$ can be determined in the same way as for the pure circularly polarized photon beam (see Eq.~\eqref{equ:Edef}). Azimuthal asymmetries caused by detection inefficiencies in $\phi$ were found to be small enough to not cause any significant false helicity asymmetries ($\ll 1$\%) due to the linear polarization component. According to \cite{Sandorfi}, $E$ is defined as
\begin{equation}
\label{equ:Edef}
E=\frac{\frac{d\sigma}{d\Omega}^{1/2} - \frac{d\sigma}{d\Omega}^{3/2}}{\frac{d\sigma}{d\Omega}^{1/2} + \frac{d\sigma}{d\Omega}^{3/2}}=\frac{N^{1/2} - N^{3/2}}{N^{1/2}+N^{3/2}} \times \frac{1}{d} \times \frac{1}{p_\gamma^{\text{circ}} p_T},
\end{equation}
where the helicity-dependent differential cross sections $\frac{d\sigma}{d\Omega}^{1/2 (3/2)}$ can be replaced by the event yields $N^{1/2}$ and $N^{3/2}$ for antiparallel or parallel combined spin configuration of the photon and proton in the initial state, respectively. The dilution factor $d$ accounts for the background from bound nuclei in the butanol data. The total systematic uncertainty comprises several aspects: the degree of circular polarization (2.7\%) that was determined through comparison of results obtained with a Mott and a M{\o}ller polarimeter, the degree of target polarization (2.8\%) based on the NMR measurements (for details see \cite{PhDthesis}), the uncertainty of the scaling factor $s$ (3.4 - 9\%) and remaining background contributions after the event selection (mostly below 2\% for the $p\pi^0$ final state and 3 - 5\% for the $n\pi^+$ final state). Furthermore, the deviations for the circular polarization degree were calculated for each coherent edge setting and an event based average deviation was calculated for the complete data set and used to correct the data. An additional 1\% systematic error was estimated for the calculation of the circular polarization degree based on comparisons between measured and calculated enhancement spectra for the coherent bremsstrahlung (see Fig.~1 in the Supplemental Material~\cite{afzal23suppl}).

Figure \ref{fig:DiaMoe} shows a comparison for the double-polarization observable $E$ between data obtained with a diamond radiator (elliptically polarized photons, blue filled points) and data obtained with an amorphous radiator (circularly polarized photons, red open points). For the data taken with the diamond radiator, the coherent edge position was chosen to be at approximately 450~MeV ($x\approx 0.3$). At this position, the maximum degree of linear polarization is relatively  high, at around $70$\%. Here, both data sets were taken during one beamtime directly one after another in time, leaving all experimental conditions, and hence all systematic effects, exactly the same. Any remaining systematic deviation should therefore be related to whether or not elliptically polarized photons can be used to determine $E$. Within the statistical precision, we observe a very good agreement between both data sets for both the $p\pi^0$ and for the $n\pi^+$ final states. This is the first experimental evidence that elliptically polarized photons can be used to measure the double-polarization observable $E$ and that Eq.~\eqref{eq:circpol} can be approximately used to determine the circular polarization degree of bremsstrahlung photons from longitudinally polarized electrons in crystals even if the coherent edge position is as low as $x\approx 0.3$. 
\begin{figure}[htb]
\includegraphics[width=\columnwidth]{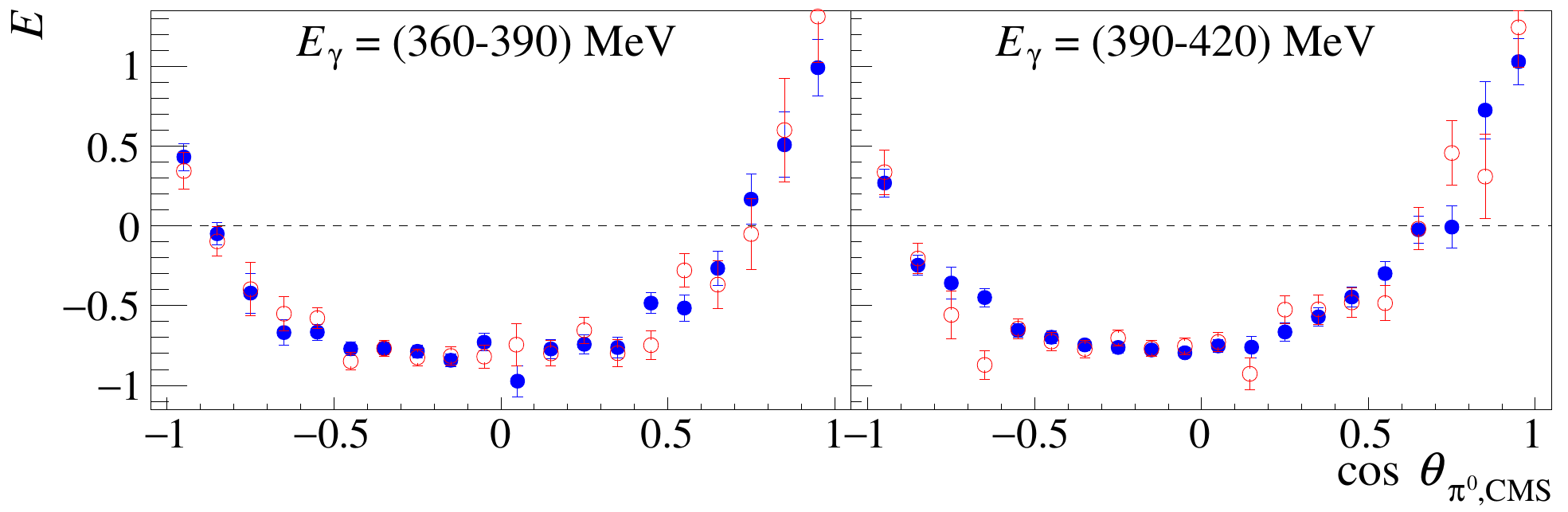}

\includegraphics[width=\columnwidth]{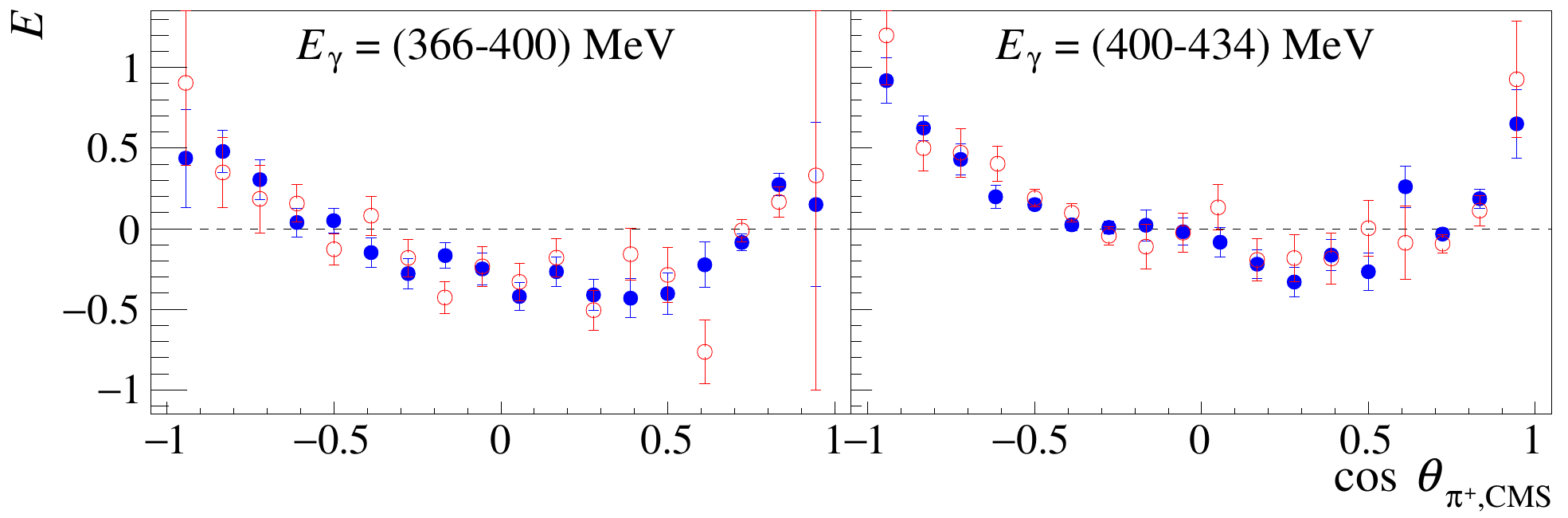}
\caption{Comparison of the results for the helicity asymmetry $E$ between measurements performed using a diamond radiator (blue filled points) and an amorphous (red open points) radiator. The upper row shows the results for the $p\pi^0$ final state and the lower row for the $n\pi^+$ final state. The shown energy bins of the diamond runs were taken with the coherent edge position at approximately 450 MeV.}
\label{fig:DiaMoe}
\end{figure}

\begin{figure}[htb]
\includegraphics[width=\columnwidth]{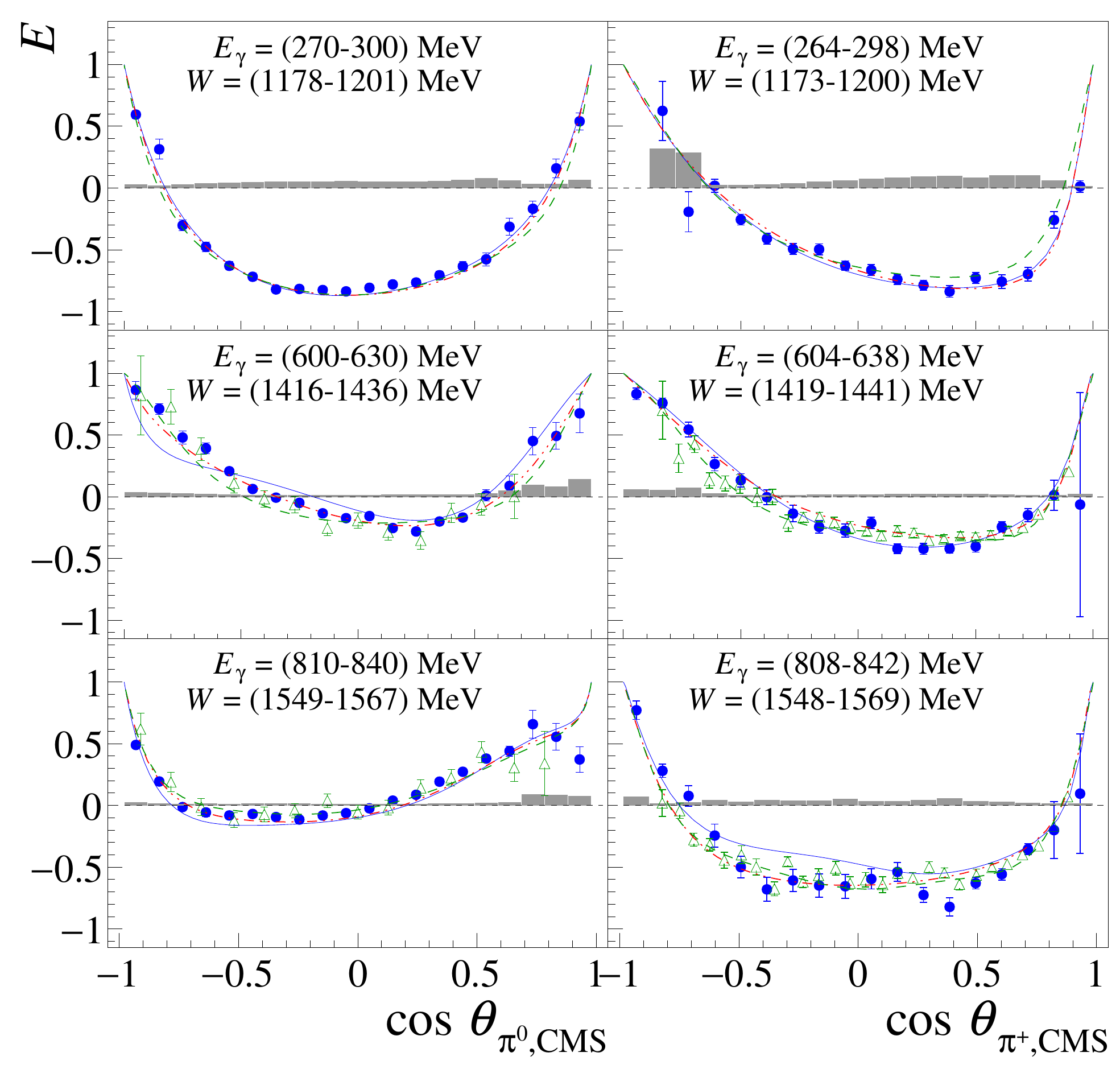}
\caption{The double-polarization observable $E$ is shown as a function of $\cos\theta$ (blue points) for three selected energy bins and for the $p\pi^0$ (left column) and the $n\pi^+$ (right column) final states. The data are compared to previous data from the CBELSA/TAPS collaboration (green, open triangles in the left column) \cite{Epi0prl,Epi0long} and from the CLAS collaboration (green, open triangles in the right column) \cite{Strauchdata}. In addition, the latest PWA are plotted as well (SAID-MA19 (blue, solid line) \cite{SAIDMA19}, BnGa (red,  dash-dotted line) \cite{BnGa2017} and J\"uBo-2017 (green, dashed line) \cite{JuBo17}). }
\label{fig:Ebsp}
\end{figure}

Selected energy bins for the double-polarization observable $E$ are depicted in Fig.~\ref{fig:Ebsp} using the full data set (all beamtimes with all diamond settings) and adding all data sets taken with an amorphous radiator. The new data for the $p\pi^0$ final state exceed the existing data in statistical precision by more than a factor of 4. In addition, the new data cover the full polar angular range that the other previous data \cite{Epi0prl,Epi0long, Strauchdata} and very recent CLAS data \cite{CLAS:2023ddn} did not achieve. In particular, new data are available for the forward angles for the $p\pi^0$ final state and backward angles for the $n\pi^+$ final state. Furthermore, it should be noted that all previous data were taken with only a circularly polarized photon beam. Thus, we emphasize that using elliptically polarized photons give the same results for the double-polarization observable $E$. 

The comparisons to the latest PWA from Scattering Analysis Interactive Dial-in (SAID) \cite{SAIDMA19}, Bonn-Gatchina (BnGa) \cite{BnGa2017} and J\"ulich-Bonn (J\"uBo) \cite{JuBo17} are displayed as well. In the energy range of the $\Delta(1232)\frac{3}{2}^+ (P_{33})$ resonance (270 MeV $\leq E_\gamma\leq$ 420 MeV), all PWA of the $p\pi^0$ final state agree very well with each other as expected since this resonance is very well known. Here, the data show an excellent agreement to the PWA within the uncertainties. Also, for the higher energies the PWA show a similar angular dependence due to the already fitted data and the new data are consistent with them. 
To show the importance of measuring $E$ for the entire polar angular range and assess the quality of the new data, a moment analysis was performed \cite{LFitPaper}. The profile function $\check{E}$, which is the product of the double-polarization observable $E$ and the differential cross section, can be expressed by a finite series of partial waves (see Eq.~\eqref{eq:LowEAssocLegParametrizationE2}), 
using an expansion of the photoproduction amplitude into electric and magnetic multipoles ${E_{\ell\pm},M_{\ell\pm}}$ and truncating it at a maximal obital angular momentum $\ell_{\text{max}}$. The angular dependence is described using associated Legendre functions $P_\ell^m (\cos\theta)$, while the energy dependence is given by a sum of bilinear products of the photoproduction multipoles, which are described by the Legendre coefficients $\left(a_{\ell_{\text{max}}}\right)^{\check{E}}_{j}$:    
{\allowdisplaybreaks
\begin{align}
\check{E}= E\cdot \frac{d\sigma}{d\Omega}_0= \frac{q}{k} \hspace*{3pt} \sum \limits_{j = 0}^{2 \ell_{\text{max}}} \left(a_{\ell_{\text{max}}}\right)_{j}^{\check{E}} \left( W \right) P_{j}^0 \left( \cos \theta \right) \mathrm{.}  \label{eq:LowEAssocLegParametrizationE2}
\end{align}
}

The BnGa PWA was utilized for the differential cross section. This choice does not lead to a bias of the extracted fit coefficients since all recent PWA (e.g.~BnGa, SAID-MA19) have included all available data sets for the differential cross section and describe them similarly well. Exchanging the BnGa PWA by the SAID PWA gave the same results for the fit coefficients within their statistical uncertainties.   
Equation \ref{eq:LegCoeff_E_4_2} shows as an example the Legendre coefficient $\left(a_{3}\right)^{\check{E}}_{2}$ in  short notation, which indicates the interference terms between different partial waves of specific orbital angular momentum $\ell$ (e.g. $S,P,D,F$ etc. correspond to $\ell=0,1,2,3$ and so on).  
{\allowdisplaybreaks
\begin{equation}
\left(a_{3}\right)^{\check{E}}_{2} = \left<P,P\right> 
+ \left<S,D\right> + \left<D,D\right>  + \left<P,F\right> + \left<F,F\right>.
\label{eq:LegCoeff_E_4_2}
\end{equation}
}
\begin{figure*}[htb]
\includegraphics[width=\textwidth]{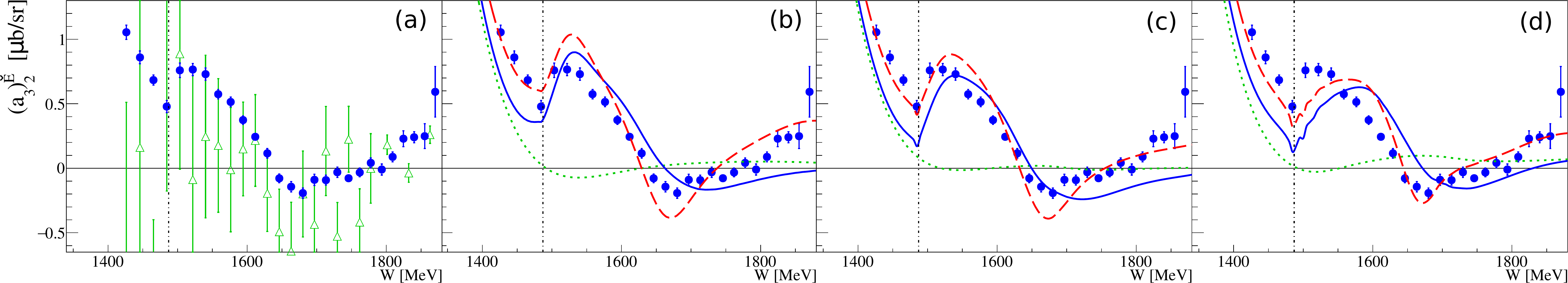}
\caption{The Legendre coefficient $\left(a_{3}\right)^{\check{E}}_{2}$ is shown as a function of the center of mass energy $W$ (blue points) for the $p\pi^0$ final state. It was extracted through fits to $\check{E}$ (see Eq.~\eqref{eq:LowEAssocLegParametrizationE2}), truncating at $\ell_{\text{max}}=3$. The results are compared in (a) to the existing CBELSA/TAPS data (green, open triangles) \cite{LFitPaper}. The continuous curves are extracted from the SAID-MA19 multipoles \cite{SAIDMA19} (b), the BnGa multipoles \cite{BnGa2017} (c) and the J\"uBo-2017 multipoles \cite{JuBo17} (d), respectively, and truncating the expansion at $\ell_{\text{max}}=1$ (green, dotted line), at $\ell_{\text{max}}=2$ (blue, solid line) and at $\ell_{\text{max}}=3$ (red, dashed line). A clear $p\eta$ cusp is visible in the new data and the PWA curves for $\ell_{\text{max}}\geq 2$ due to the $\left<S,D\right>$ interference term.}
\label{fig:LegCoeff2}
\end{figure*}
Figure \ref{fig:LegCoeff2} shows the energy dependence of the coefficient $\left(a_{3}\right)^{\check{E}}_{2}$ for the $p\pi^0$ final state. It was obtained by fitting all energy bins using Eq.~\eqref{eq:LowEAssocLegParametrizationE2} and truncating at $\ell_{\text{max}}=3$, which is sufficient to get a good $\chi^2/\text{ndf}$ value for the fit. Truncating at $\ell_{\text{max}}=4$ does not change the fit results and the last two fit coefficients are consistent with zero. The data exhibit a sudden change in the slope at precisely the $p\eta$ photoproduction threshold (at $W=1487$ MeV). The opening of the $p\eta$ final state shows up as a cusp in the $S$ wave of the $p\pi^0$ final state and is visible in the coefficient $\left(a_{3}\right)^{\check{E}}_{2}$ due to the $\left<S,D\right>$ interference term (see Eq.~\eqref{eq:LegCoeff_E_4_2}). This claim is further confirmed by comparing the data to the different PWA curves, which were extracted using the PWA multipoles and truncating the expansion at different $\ell_{\text{max}}$. The blue solid curves in Fig.~\ref{fig:LegCoeff2} contain the $\left<S,D\right>$ interference term and clearly show the $p\eta$ cusp  for all three PWAs. In contrast to the new data, the previous CBELSA/TAPS data \cite{Epi0prl, Epi0long} lack the precision required to observe the $p\eta$ cusp. This is not just a result of lower statistical precision, but also a consequence of not having measured the double-polarization observable $E$ for the entire polar angular range (see Fig.~\ref{fig:Ebsp}). The comparison between the different PWA solutions reveals existing discrepancies in the description of the $p\eta$ cusp in their approaches, which was also noted by Anisovich et al.~\cite{Anisovich:2016vzt}. The new data will help to improve the description of the $p\pi^0$ $S$ wave energy dependence, which is relevant for the $S$-wave resonances ($J^P=\frac{1}{2}^-$) $N(1535), \Delta(1620), N(1650), N(1895)$, and $\Delta(1900)$, and through interference with other waves can also affect all other partial waves and resonances. 

In summary, the double-polarization observable $E$ was measured for the first time using elliptically polarized photons and a longitudinally polarized butanol target for both the $p\pi^0$ and the $n\pi^+$ final states. The results show a very good agreement between data that were taken with only a circularly polarized photon beam, providing the possibility of measuring polarization observables that require either linearly or circularly polarized photons at the same time and thus enabling a more time and cost efficient, as well as self-consistent measurement (see table~1 in the Supplemental Material~\cite{afzal23suppl}). In addition, we provide a first calculation for the circular polarization degree when using a diamond radiator and show that only small deviations occur for the circular polarization degree. The new data have a high precision and a large angular coverage, which leads to a precise observation of the $p\eta$ cusp in the $p\pi^0$ final state for the first time in a double-polarization observable. The correct implementation of pronounced cusp-effects is important for the correct extraction of the amplitudes and resonance parameters in a PWA. Here, the new data will help to resolve the observed discrepancies between the different PWA solutions at and above the $p\eta$ threshold energy. The new measurement method presented may prove valuable in other fields besides light baryon spectroscopy, e.g. for the measurement of the nucleon polarizabilities in real Compton scattering \cite{A2:2019bqm,A2CollaborationatMAMI:2021vfy}.

We thank Prof. Richard T. Jones for the fruitful discussions regarding the calculation of the circular polarization degree. We wish to further acknowledge the outstanding support of the accelerator group and  operators of MAMI. This work was supported by Schweizerischer Nationalfonds (200020-156983, 132799, 121781, 117601), Deutsche Forschungsgemeinschaft (SFB443,  SFB1044, SFB/TR16), the INFN-Italy, the European Community-Research Infrastructure Activity under FP7 programme (Hadron Physics, grant agreement No. 227431), the U.~K.~Science and Technology Facilities Council grants ST/J000175/1, ST/G008604/1, ST/G008582/1, ST/J00006X/1, ST/V001035/1 ST/V002570/1, ST/P004385/2, ST/T002077/1, and ST/L00478X/2, the Natural Sciences and Engineering Research Council (NSERC, FRN: SAPPJ-2015-00023), Canada. This material is based upon work also supported by the U.S. Department of Energy, Office of Science, Office of Nuclear Physics Research Division, under Award Numbers DE-FG02-99-ER41110, DE-FG02-88ER40415, DE-FG02-01-ER41194, DE-SC0014323 and DE-SC0016583 and by the National Science Foundation, under Grant Nos. PHY-1039130 andIIA-1358175.

\bibliographystyle{apsrev4-2}
\bibliography{E-prl.bib}

\end{document}